\documentclass[prb,twocolumn]{revtex4-1}
\usepackage{amsmath}
\usepackage[dvips]{psfrag,graphicx}
\usepackage{verbatim}
\usepackage[usenames]{xcolor}
\usepackage{subfigure}
\usepackage[colorlinks=true, pdfstartview=FitV, linkcolor=blue, citecolor=violet, urlcolor=blue]{hyperref}
\usepackage{float}

\graphicspath{{../Figures/}{./}} 
\begin{document}
%



\setcounter{MaxMatrixCols}{20}

\title{Zero energy modes in a superconductor with ferromagnetic adatom chains and quantum phase transitions}

\author{Tilen \v Cade\v z$^{1,2,3}$}
\email{tilen.cadez@ijs.si}
\author{Pedro D. Sacramento$^{2,1}$}
\affiliation{ $^{1}$ Beijing Computational Science Research Center, Zhongguancun Software Park II, No. 10 West Dongbeiwang Road, Haidian District, Beijing, 100094, China}
\affiliation{ $^{2}$ CeFEMA, Instituto Superior T\'{e}cnico, Universidade de Lisboa, Av. Rovisco Pais, 1049-001 Lisboa, Portugal}
\affiliation{ $^{3}$ Jo\v zef Stefan Institute, 1000 Ljubljana, Slovenia}

\date{\today}



\begin{abstract}
We study Majorana zero energy modes (MZEM) that occur in a s-wave superconducting surface, at the ends of a ferromagnetic (FM) chain of adatoms, in the presence of Rashba spin-orbit interaction (SOI) considering both non self-consistent and self-consistent superconducting order. We find that in the self-consistent solution the average superconducting gap function over the adatom sites has a discontinuous drop with increasing exchange interaction at the same critical value where the topological phase transition occurs. We also study the MZEM for both treatments of superconducting order and find that the decay length is a linear function of the exchange coupling strength, chemical potential and superconducting order. For wider FM chains the MZEM occur at smaller exchange couplings and the slope of the decay length as a function of exchange coupling grows with chain width. Thus we suggest experimental detection of different delocalization of MZEM in chains of varying widths. We discuss similarities and differences between the MZEM for the two treatments of the superconducting order.
\end{abstract}

\maketitle
\section{Introduction}

Non-magnetic impurities in a conventional superconductor have no pair-breaking effect and do not lead to qualitative change~\cite{Anderson59}, in contrast to the case of unconventional gapless superconductors where their effect is pair-breaking~\cite{Ueda85}. On the other hand magnetic impurities in a conventional superconductor have interesting effects~\cite{Balatsky06}. A single magnetic impurity coupled to the conduction electron spin density gives rise to a local bound state, known as Yu-Shiba-Rusinov (YSR) state, that is created by breaking a Cooper pair and capturing an electron with the appropriate spin component~\cite{Yu65, Shiba68, Rusinov69, Sakurai70, Salkola97, Flatte97}. As the exchange coupling $J$ increases, a first order quantum phase transition (QPT) occurs, at which point the gap function has a $\pi$ shift and the magnetization of the conduction electrons jumps from zero to $1/2$. This phase transition has been detected experimentally~\cite{Yazdani97, Ji08, Hatter15} and is signalled in various quantities including quantum information signatures~\cite{Sacramento07b, Paunkovic08}. As it turns out~\cite{Yu65, Shiba68, Rusinov69} the YSR states come in pairs that lay inside the gap (one at positive and one at negative energy) and tend to lower energies as the coupling grows. At the quantum critical point there is a level crossing, such that the bound state has a small but finite energy. Increasing the number of impurities, more states appear inside the gap (two per impurity) and as the coupling increases, a series of quantum phase transitions occurs and the magnetization of the conduction electrons changes in increasing plateaus. Interesting interference phenomena occur as the number of impurities grows and extended states appear along the chain of impurities placed on the two-dimensional superconductor~\cite{Ji08, Morr03, Morr04, Morr06, Sacramento07a}.

A quantum treatment of impurities with spin 1/2 describes the competition between the screening of the impurity spin and the superconducting correlations, which lead to a similar type of QPT between the spin-doublet and spin-singlet state for the unscreened and screened impurity spin, respectively~\cite{Zittartz70, Satori92, Sakai93}. Real magnetic impurities however, have higher spin, where magnetic anisotropy plays an important role~\cite{Otte08}. The effects of magnetic anysotropy on the subgap excitations induced by quantum impurities in conventional superconductor~\cite{Zitko11} were recently measured~\cite{Hatter15} in Manganese phthalocyanine (MnPc) on top of superconductor Pb, where Mn has high spin 3/2. Different possible ways of adsorption of MnPc molecules enabled the precise spectroscopy of YSR states across the QPT using scanning tunnelling microscope (STM).

Increasing the number of magnetic impurities eventually leads to destruction of superconductivity. If the magnetic impurities have arbitrary orientations and locations on the superconductor the pair breaking effect leads to gapless superconductivity and eventually destruction of superconductivity~\cite{Abrikosov61} for small concentration of impurities of the order of a few percent. A higher robustness of superconductivity to the increasing number of magnetic impurities has been found if they are correlated, particularly if their locations are not random but organized in some patterns~\cite{Sacramento10}. These regular distributions allow quite high concentrations of impurities without destruction of superconductivity.

In 2014 an experiment was reported~\cite{Nadj-Perge14} in a system of a chain of magnetic adatoms (Fe) placed on top of a two-dimensional conventional superconductor (Pb). Localized zero energy modes were detected at the edges of adatom chain using STM, which seems to provide evidence of MZEM. These zero energy modes have attracted considerable attention recently~\cite{Hasan10, Qi11, Leijnse12, Beenakker13, Beenakker15} because of their non-abelian statistics, which can be used for quantum information storage and manipulation~\cite{Kitaev01, Kitaev03, Nayak08, Alicea12, DasSarma15}. A key model exhibiting such topological properties is the one-dimensional Kitaev model of spinless fermions with p-wave superconducting pairing~\cite{Kitaev01}. More feasible models based on a conventional s-wave superconductor in proximity to either topological insulator~\cite{Fu08} or semiconducting nanowires in the presence of a Zeeman field~\cite{Lutchyn10, Oreg10} greatly enhanced the interest in the field. The later proposal appears to be experimentaly realized in a zero-bias conductance peak measurements in a semiconducting nanowire on top of a conventional superconductor~\cite{Mourik12, Finck13}. Although all these experiments support the existence of MZEM they are for now inconclusive~\cite{Finck13, Lee14, Peng15, Dumitrescu15}.

In the experiment~\cite{Nadj-Perge14}, the zero energy states are sharply located at the edges. Complementing the experiment was a theoretical study of the system, which in its simplest form considers 2D superconducting surface with nearest neighbour hopping, chemical potential, exchange interaction with the ferromagnetically aligned magnetic adatoms, Rashba SOI and proximity induced s-wave superconductivity. As parameters of the model change, the system evolves from trivial to topological phase, as determined by the non self-consistent solution of the problem, justifiable by a proximity induced superconducting order or by neglecting the change of the gap function and consequently of other physical quantities in the vicinity of the magnetic adatoms.

Moreover, there are many theoretical studies showing that regularly positioned chains of adatoms give rise to MZEM. These were given for the cases of random~\cite{Choy11}, spiral~\cite{Kjaergaard12, Martin12, Nadj-Perge13, Nakosai13, Braunecker13, Klinovaja13, Vazifeh13, Pientka13, Pientka14, Poyhonen14, Rontynen14, Reis14, Kim14, Li16, Hoffman16}, AFM~\cite{Heimes14, Heimes15} and FM~\cite{Nadj-Perge14, Li14b, Heimes15,  Brydon15, Hui15, Peng15, Poyhonen16} orderings of the magnetic adatoms. A stable spiral magnetic structure of the 1D adatom chain is established via the Ruderman-Kittel-Kasuya-Yosida (RKKY) interaction~\cite{Ruderman54, Kasuya56, Yosida57} mediated by the electrons in the underlaying superconductor, which has to be treated fully self-consistently~\cite{Braunecker13,  Klinovaja13, Vazifeh13}. A more realistic 2D model of a superconducting surface coupled to a 1D chain of magnetic moments shows~\cite{Reis14, Heimes15}, that besides spiral ordering, there are regions in the phase diagram where the magnetic moments arrange either FM or AFM. These however, can in the presence of SOI, still host MZEM which occur in multiple topological phases~\cite{Heimes15, Poyhonen16} even in the presence of periodic modulation of superconducting order parameter~\cite{Hoffman16}. Recently magnetic chains and the competition between MZEM with the zero energy modes at the border of a triplet superconductor were also studied~\cite{Sacramento15}. The importance and effects of the self-consistent solution on the order parameter in triplet superconductor were also discussed very recently~\cite{Mercaldo16}.

In this work we are interested in comparing the results obtained from the non self-consistent solution with those obtained by treating the superconducting gap function self-consistently. The case of a few impurities has shown the relevance of a full treatment~\cite{Sacramento07a, Sacramento10} and we aim to see how the results change as we pursue a more detailed solution. As was already shown previously for spiral magnetic chains~\cite{Nadj-Perge13, Reis14} qualitatevely both treatments give similar results and here we confirm such results for the case of FM chain in the presence of Rashba SOI, with notable quantitative shift of the regions with MZEM in chemical potential. We study also Majorana decay lengths and find that as a function of exchange interaction, chemical potential and the superconducting gap function (effective attractive interaction for the self-consistent superconductivity) the decay length is a linear function with positive, zero and negative slope, respectively. By considering FM chains of various widths, we show that the wider chains correspond to a narrow chain with increased effective exchange interaction. Due to the linear dependence of the Majorana decay length on exchange interaction different delocalizations of MZEM in the STM measurements should be detectable on chains of various widths.

An important difference between the two treatments of superconductivity is that in the self-consistent not only topological phase transition but also a series of quantum phase transitions occur. The later are due to various ingap level crossings and can be observed in changes of total magnetization of the electron spin density, local spin density, local density of states, various quantum information measures and the local order parameter. In this work we focus on the later and show that the local superconducting gap on the adatom sites experiences a discontinuous drop at the exchange coupling where the topological phase transition occurs.

The paper is organized as follows: in section \ref{sec:II} we introduce the considered model. In section \ref{sec:III} we present solutions of the model for both treatments of the superconducting order parameter. We study energy levels in subsection \ref{sec:IIIA}, discuss topological invariant in subsection \ref{sec:IIIB} study the MZEM wavefunctions in subsection \ref{sec:IIIC} and consider the case of wider chains and the Majorana decay lengths in subsection \ref{sec:IIID}. In section \ref{sec:IV} we focus on the self-consistent treatment of superconductivity and study local order parameter. We present our conclusions in section \ref{sec:V}.

\section{Model}\label{sec:II}

\begin{figure}[bht]
\centering
\includegraphics[width=0.3\textwidth]{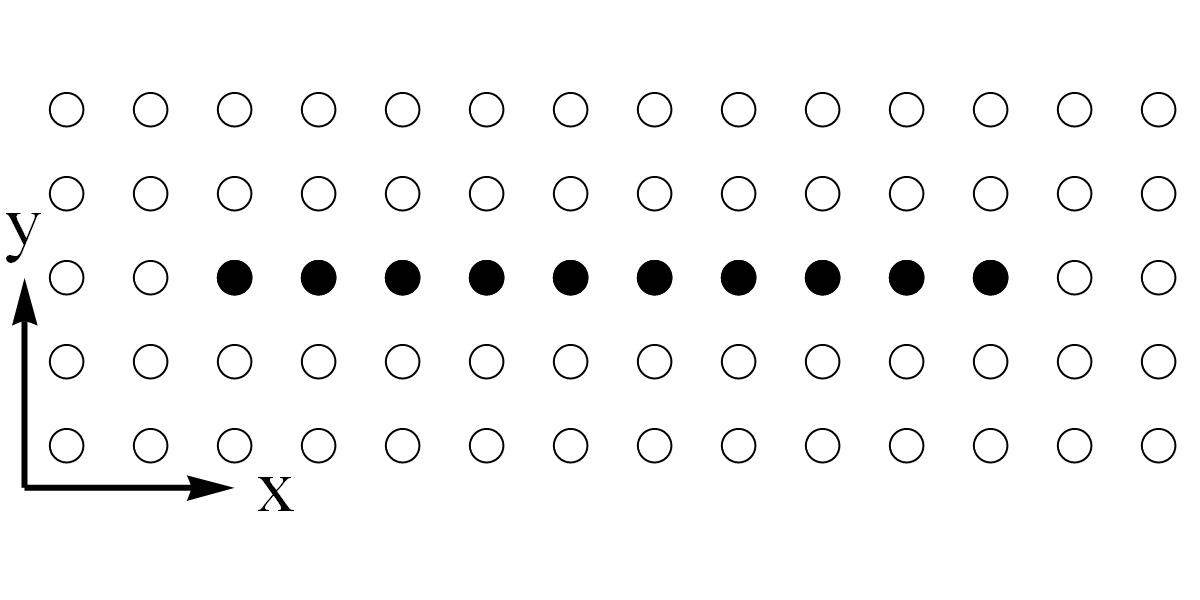}
\caption{Systems considered: Example of a 2D square lattice of size $N_x \times N_y$, with the magnetic adatom chain (black) of length $L$ and width $w = 1$.}\label{fig:2.1}
\end{figure}

We consider a 2D s-wave superconducting surface with adsorbed magnetic adatoms, which we describe as classical spins and assume their FM alignment. The classical spins are placed along the line of length $L$ and width $w$ and oriented along the normal to the surface. The SC surface is a 2D square lattice as shown in Fig.\ref{fig:2.1} and the Hamiltonian of the system considered is
\begin{eqnarray}\label{H}
H &=& -t \sum_{\langle i,j \rangle, s} c_{is}^{\dagger} c_{js} - \mu \sum_{i,s} c_{is}^{\dagger} c_{is} \\
&+& i \alpha \sum_{i,s,s'} \bigl( c_{i+\delta x,s}^{\dagger} \sigma_{s s'}^y c_{i, s'} - c_{i+\delta y,s}^{\dagger} \sigma_{s s'}^x c_{i, s'} + {\it h.c.}\bigr) \nonumber \\
&+& \sum_i \bigl(\Delta_i c_{i \uparrow}^{\dagger} c_{i \downarrow}^{\dagger} + \Delta_i^* c_{i \downarrow} c_{i \uparrow} \bigr) - \sum_{i,s,s'} J_i c_{i,s}^{\dagger} \sigma_{s s'}^z c_{i, s'}, \nonumber
\end{eqnarray}
where the first term describes the electron hopping between two adjacent lattice sites, the second term is the chemical potential $\mu$, the third is the Rashba SOI, forth term is the superconducting s-wave pairing, with site dependent order parameter $\Delta_i$ and the last term is exchange interaction with the magnetic adatoms located at specific sites. We consider $J_i = J$ at these adatom sites. $\delta x (\delta y)$ are vectors to the next neighbouring sides along x (y) direction, $\sigma^a$ are Pauli spin matrices with $a = x,y,z$ and $s,s' = \uparrow, \downarrow$. Note that the indices $i,j = 1, \ldots, N$ , where $N$ is the total number of sites. 

We diagonalize the Hamiltonian using the Bogoliubov transformation
\begin{eqnarray}\label{Bogoliubov}
c_{i \uparrow} = \sum_n \bigl[ u_n(i, \uparrow) d_n - v_n^*(i, \uparrow) d_n^{\dagger}\bigr], \nonumber \\
c_{i \downarrow} = \sum_n \bigl[ u_n(i, \downarrow) d_n + v_n^*(i, \downarrow) d_n^{\dagger}\bigr],
\end{eqnarray}
where $n$ is the complete set of states, $u_n$ and $v_n$ are related to the eigenfunctions of the Hamiltonian \eqref{H} and $d_n^{\dagger} (d_n)$ are the new fermionic quasiparticle creation (annihilation) operators. We calculate the coefficients $u_n(i, s)$ and $v_n(i, s)$ by solving Bogoliubov-de Gennes (BdG) equations~\cite{deGennes}, which can be written as 
\begin{eqnarray}
{\cal{H}} \psi_n = \varepsilon_n \psi_n,
\end{eqnarray}
with the $4 N$ dimensional vector 
\begin{eqnarray}
\psi_n(i) = \begin{pmatrix}
u_n(i, \uparrow) \\
v_n(i, \downarrow) \\
u_n(i, \downarrow) \\
v_n(i, \uparrow) \\
\end{pmatrix}, \nonumber
\end{eqnarray}
and the matrix ${\cal{H}}$ at site $i$ is
\begin{widetext}
\begin{eqnarray}\label{calH}
{\cal{H}} = \begin{pmatrix}
  - t (\eta_x^{+} + \eta_y^{+}) - \mu - J_{i} & \Delta_i & \alpha (-\eta_x^- + i \eta_y^-) & 0 \\
  \Delta_i^* &  t (\eta_x^{+} + \eta_y^{+}) + \mu - J_{i} & 0 & \alpha (\eta_x^- - i \eta_y^-) \\
  \alpha (\eta_x^- + i \eta_y^-) & 0 &  - t (\eta_x^{+} + \eta_y^{+}) - \mu + J_{i} & \Delta_i \\
  0 & - \alpha (\eta_x^- + i \eta_y^-) & \Delta_i^* &  t (\eta_x^{+} + \eta_y^{+}) + \mu + J_{i} \\
\end{pmatrix}, 
\end{eqnarray}
\end{widetext}
where $\eta_l^{\pm} f(i) = f(i + l) \pm f(i - l)$, for $l = x,y$, {\it{i.e.}} displacement along $x$ and $y$ and coefficients $f(i)$ are the particle and hole coefficients $u_n(i,s), v_n(i,s)$. Unless stated otherwise we set the hopping to unity, {\it i.e.} $t = 1$ and express all the energies in this unit.

We solve BdG equations by diagonalizing a $(4 N) \times (4 N)$ matrix, with the gap either constant, which we refer to as non self-consistent solution or calculated self-consistently, imposing at each iteration that
\begin{equation}\label{Di1}
\Delta_i = V (\langle c_{i \uparrow} c_{i \downarrow} \rangle - \langle c_{i \downarrow}  c_{i \uparrow}  \rangle)/2,
\end{equation}
where $V$ is an effective attractive interaction between the electrons. The superconducting gap function is calculated via
\begin{equation}\label{Di2}
\Delta_i = V \sum_{n, \varepsilon_n < \omega_D} \bigl( 1/2 - f_n \bigr) \bigl[ u_n(i, \uparrow) v_n^*(i, \downarrow) + u_n(i, \downarrow) v_n^* (i, \uparrow) \bigr],
\end{equation}
where $\omega_D$ is Debye frequency, the sum runs over all the positive energy eigenstates with values smaller than $\omega_D$, $f_n = 1/[e^{\varepsilon_n/T} + 1]$ is the Fermi function and $T$ is the temperature. In this work we set $\omega_D = 2$, $T = 0.001$ and find that calculated $\Delta_{i}$ has in all cases considered negligible imaginary part. The dimension of the systems considered are $N_x \ge L$ where the chain length $L$ is up to 180 sites and the lateral dimension $N_y$ is between 11 and 31.

\section{Majorana zero energy modes}\label{sec:III}
\subsection{Energy levels}\label{sec:IIIA}

First let us focus on the single-particle energy levels of eq. \eqref{H} for self-consistent treatment of superconductivity. In the absence of magnetic adatoms ($J = 0$) and SOI ($\alpha = 0$), the energy states in a clean homogeneous s-wave superconductor are gapped, with the gap value of $\Delta$. By adding a single magnetic impurity, two YSR local bound states appear within the gap~\cite{Yu65, Shiba68, Rusinov69}, one at positive and one at a symmetric negative energy. If the exchange coupling $J$, between the impurity spin and the conduction electron spin density is weak enough, the problem may be treated classically~\cite{Satori92} and the impurity spin acts as a local magnetic field that interacts with the superconductor through the Zeeman effect. In this work we consider such case. Note also that $J$ is actually a product of the exchange coupling and the magnitude of the impurity spin $S$. As the exchange coupling $J$ increases, the values of the two levels drop towards zero until a critical value of $J$ is reached, where a QPT occurs. With the further increase of $J$ the levels start to repel and eventually merge into the gapped states. 

In the case of $N_i$ impurities, there are $2 N_i$ YSR states inside the gap for small $J$. Similar to the case of a single impurity, the energy values of these states approach zero by increasing the exchange coupling and in the case of an ordered array they form minibands~\cite{Sacramento07a}. By increasing the exchange coupling several level crossings occur (all at finite energy) accompanied by a series of first order QPTs. 

\begin{figure*}[bht]
\centering
\includegraphics[width=0.75\textwidth]{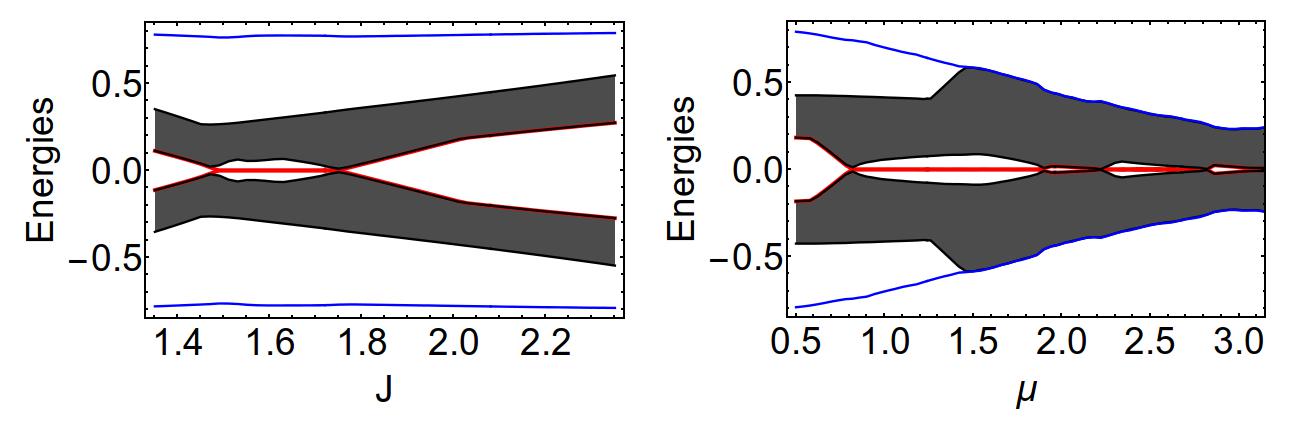}
\caption{Energies of the lowest $2 (L+1)$ single-particle energy states as a function of exchange interaction $J$ and the chemical potential $\mu$ for the self-consistent superconductivity. Lowest energy level is shown with full red line. The second and $L$-th single-particle energy levels are given with full black line and define the minigap in the areas where the lowest level has zero energy. The YSR states, except the lowest one, are shown in the shaded area, which defines the ingap miniband. The $L+1$-th state is given by blue line and when well separated from the $L$-th state, shows the behaviour of the superconducting gap. The parameters used are $L = 90$, $N_y = 11$, $\mu = 0.6$, $J = 2$, $\alpha = 0.3$ and $V = 4$.}\label{fig:3.1}
\end{figure*}

In Fig.~\ref{fig:3.1} are shown the lowest $N_i + 1$ energies as a function of exchange coupling for the chain of length $L = N_i$ and width $w = 1$, where the local spins are FM aligned. Increasing the exchange coupling from low value a YSR miniband approaches zero energy and at some critical exchange coupling the lowest state becomes practically zero. By increasing the exchange coupling the lowest energy state remains zero, while the energy of the first excited state increases to a value of about 1/10 of the superconducting gap. By further increasing $J$ the first excited state, together with the whole miniband returns to the zero energy state and the minigap closes. Still increasing the exchange coupling the whole miniband moves toward the superconducting gap, which has all along remained nearly constant, as will be discussed in more detail below. Also shown in Fig.~\ref{fig:3.1} is similar behaviour of the lowest $L + 1$ energies at fixed exchange coupling as a function of chemical potential. There are several minigap closings and reopenings and two regions with the lowest energy state remaining in vicinity of zero, while the minigap increases. At fixed exchange interaction $J$ the superconducting gap reduces with increasing chemical potential. We also note that the superconducting gap function is in the case of self-consistent superconductivity site-dependent and changes its average value by changing either chemical potential, SOI or effective attractive interaction. Similar effects were already observed in various magnetic and superconducting hybrid systems~\cite{Yang04, Gillijns05, Stamopoulos05} and studied theoretically in magnetic chains on top of a superconducting surface~\cite{Sacramento07a, Reis14} in the absence of SOI.

\begin{figure*}[bht]
\centering
\includegraphics[width=0.75\textwidth]{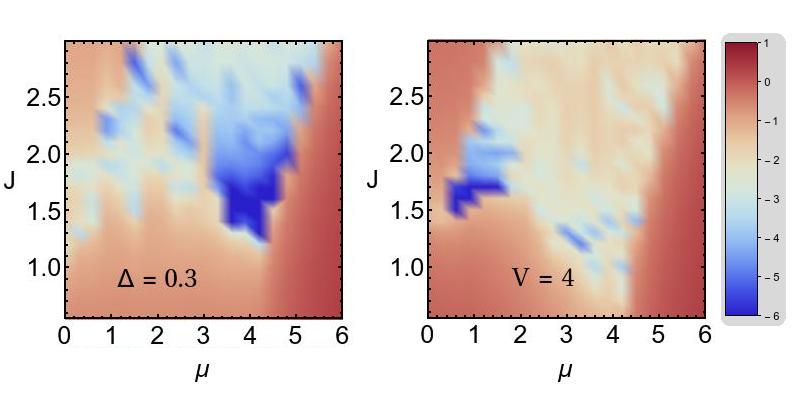}
\caption{Contour diagrams of the lowest single-particle state energy as functions of chemical potential $\mu$ and exchange coupling $J$ for a non-self consistent and self-consistent treatments are shown in left and right panel, respectively. Note the logarithmic scale used, with the cutoff at exponent $-6$. The parameters used are $L = 90$, $N_y = 11$ and $\alpha = 0.3$.}\label{fig:3.2}
\end{figure*}
%

The results for non self-consistent treatment of superconductivity are qualitatively similar for lowest energy states, but quite different when considering lowest $L+1$ energy levels. The lowest $L$ energy states are within the superconducting gap $\Delta$ only in vicinity of zero chemical potential. Away from zero chemical potential no clear superconducting gap is observed.

Next we focus on the lowest single-particle state energy. Examples of contour plots as a function of chemical potential $\mu$ and exchange coupling $J$ for fixed values of superconducting gap $\Delta$ (effective attractive interaction $V$) and SOI $\alpha$ are presented in Fig.~\ref{fig:3.2} for non self-consistent (self-consistent) superconductivity. Note that the energy scale used is logarithmic, with the cutoff at the energy $10^{-6}$. Blue (red) areas correspond to lower (higher) energies. The low energy area is mainly concentrated in the vicinity of the chemical potential $\mu \sim 4$, $\mu \sim 1$ for non self-consistent, self-consistent superconductivity, respectively. In both cases the low energy area appears at some minimal exchange coupling, which increases with increasing superconducting gap (effective attractive interaction $V$) in the case of non self-consistent (self-consistent) treatment of superconductivity.

\subsection{Topological invariant}\label{sec:IIIB}
To characterize the zero energy states we calculate the Majorana number~\cite{Kitaev01} $M$, which shows the presence ($M = -1$) or absence ($M = 1$) of MZEM at the edges of chains in the system. For translationally invariant system the Majorana number is given as
\begin{eqnarray}\label{M}
M = {\mathrm{Sign}} \Bigl[ {\mathrm{Pf}} \bigl( A_p \bigr) {\mathrm{Pf}} \bigl( A_a \bigr) \Bigr]
\end{eqnarray}
where Pf($A$) denotes the Pfaffian of a matrix $A$ and $A_p$ ($A_a$) is the Hamiltonian with periodic (anti-periodic) boundary conditions, rewritten in Majorana basis. Introducing Majorana fermion operators $a_{2 i - 1, s} = c_{i,s} + c_{i,s}^{\dagger}$, $a_{2 i, s} = i \bigl(c_{i,s} - c_{i,s}^{\dagger}\bigr)$, for which holds $a_{i,s}^{\dagger} = a_{i,s}$ and $a_{i,s} a_{j,s'} + a_{j,s'} a_{i,s} = 2 \delta_{i,j} \delta_{s,s'}$ the Hamiltonian \eqref{H} is 
\begin{widetext}
\begin{eqnarray}\label{HM}
H &=& - i/4 \sum_{i} \Bigl\{ \mu \sum_{s} a_{2 i - 1, s} a_{2 i, s}
+ t \sum_{s, \delta = \delta x, \delta y} \bigl[ a_{2 i - 1, s} a_{2 (i + \delta), s} + a_{2 (i + \delta) - 1, s} a_{2 i, s} \bigr] - \nonumber \\
&-& \alpha \bigl[ a_{2 (i + \delta x) - 1, \uparrow} a_{2 i, \downarrow} + a_{2 i - 1, \downarrow} a_{2 (i + \delta x), \uparrow} - a_{2 (i + \delta x) - 1, \downarrow} a_{2 i, \uparrow} - a_{2 i - 1, \uparrow} a_{2 (i + \delta x), \downarrow} - \nonumber \\ 
&-& a_{2 (i + \delta y) - 1, \uparrow} a_{2 i - 1, \downarrow} - a_{2 (i + \delta y) - 1, \downarrow} a_{2 i - 1, \uparrow} - a_{2 (i + \delta y), \uparrow} a_{2 i, \downarrow} - a_{2 (i + \delta y), \downarrow} a_{2 i, \uparrow} \bigr] + \nonumber \\
&+& {\mathrm{Re}}(\Delta_i) \bigl[ a_{2 i - 1, \uparrow} a_{2 i, \downarrow} - a_{2 i - 1, \downarrow} a_{2 i, \uparrow} \bigr] + {\mathrm{Im}}(\Delta_i) \bigl[ a_{2 i, \uparrow} a_{2 i, \downarrow} - a_{2 i - 1, \uparrow} a_{2 i - 1, \downarrow} \bigr] + \nonumber \\
&+& J_i \bigl[ a_{2 i - 1, \downarrow} a_{2 i, \downarrow} - a_{2 i - 1, \uparrow} a_{2 i, \uparrow} \bigr] \Bigr\}.
\end{eqnarray}
\end{widetext}

Let us first consider a 1D chain ($N_y = 1$), which is just the discrete case of a single-channel quantum wire~\cite{Lutchyn10, Oreg10}. Taking $\Delta$ as homogeneous and using the Fourier transform $a_{2 j - 1, s} = 1/\sqrt{N} \sum_k {\mathrm{Exp}}[- i k j] a_{k,1,s}$ and $a_{2 j, s} = 1/\sqrt{N} \sum_{k} {\mathrm{Exp}}[- i k j] a_{k,2,s}$ the Hamiltonian \eqref{HM} can be rewritten in the basis $a_k = (a_{k,1,\uparrow}, a_{k,1,\downarrow}, a_{-k,2,\uparrow}, a_{-k,2,\downarrow})^T$ as
\begin{eqnarray}
H = \frac{i}{4} \sum_k a_k^{\dagger} A(k) a_k,
\end{eqnarray}
where a skew-symmetric matrix $A(k)$ has the following nonzero elements
\begin{eqnarray}
A_{1,3}(k) = \varepsilon_k + J, \quad \phantom{-} A_{1,4}(k) = \Delta + 2 i \alpha \sin(k), \nonumber \\
A_{2,3}(k) = - \Delta - 2 i \alpha \sin(k) , \quad A_{2,4}(k) = \varepsilon_k - J,
\end{eqnarray}
with $\varepsilon_k = - \mu - 2 t \cos(k)$. The Majorana number can be calculated using eq. \eqref{M}, where $A_p = A(0)$ and $A_a = A(\pi)$, thus we find $M = - 1$ for 
\begin{eqnarray}
\sqrt{\Delta^2 + (\mu - 2 t)^2} < J < \sqrt{\Delta^2 + (\mu + 2 t)^2}.
\end{eqnarray}
Note that the topologically non-trivial region is for 1D case independent on the SOI. As is well known~\cite{Martin12}, the 1D chain in constant magnetic field (here $J$) and SOI is equivalent to a 1D chain in spiral magnetic field, which is evident also by comparing the matrix $A(k)$ to the one given in Ref.~\cite{Nadj-Perge13}. 

\begin{figure*}[htb]
\centering
\includegraphics[width = 0.75\textwidth]{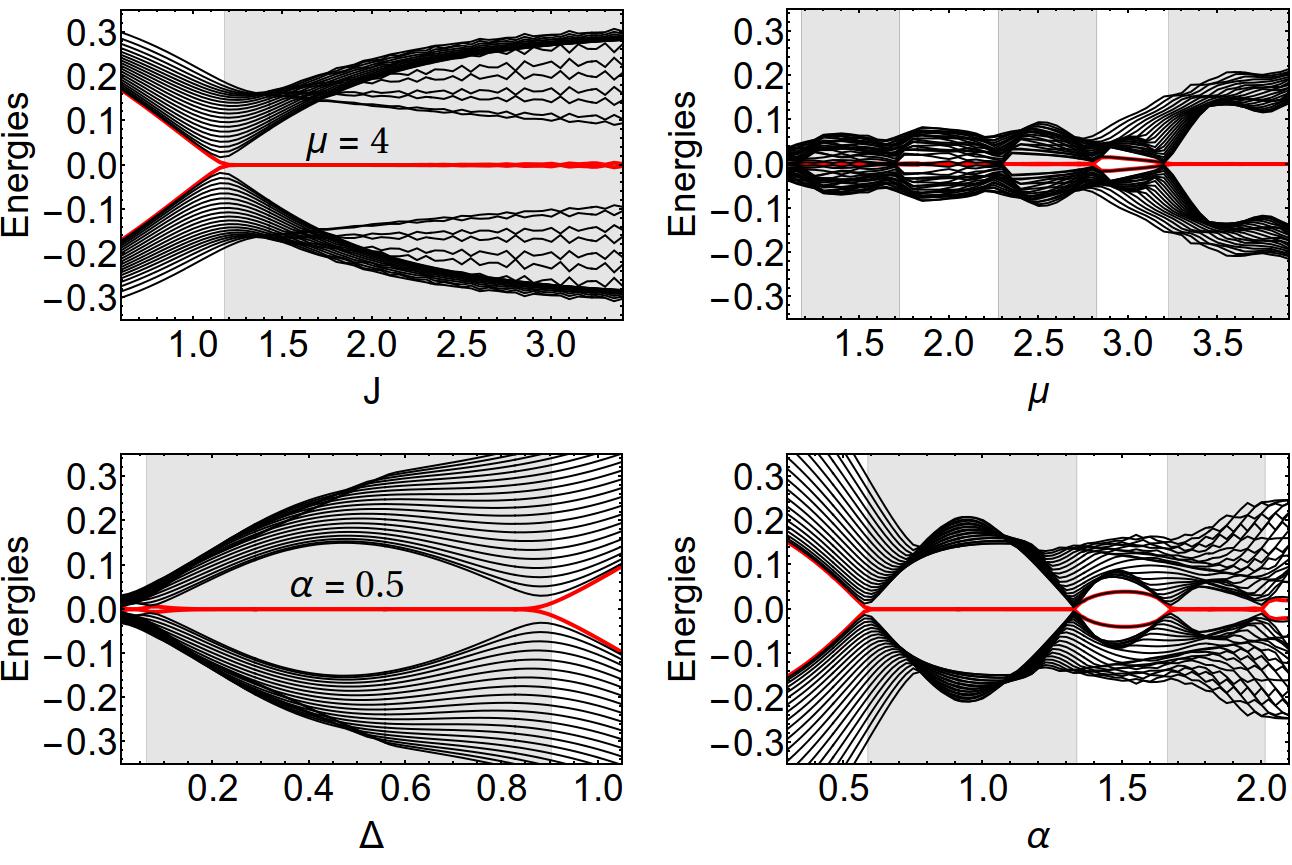}
\caption{The 40 lowest energy states as a function of model parameters: chemical potential $\mu$ , exchange coupling $J$, order parameter $\Delta$ and spin-orbit interaction $\alpha$. Unless given in the panels, the parameters used are $L =N_x = 90,$ $N_y = 11,$ $\mu = 5, J = 2, \alpha = 0.3$ and $\Delta = 0.3$. Shaded area denotes topological regions where $M = -1$ and the localized MZEM occur at the edges of adatom chain. A slight disagreement at the gap closings as a function of $\Delta$ is a finite size effect, which we confirmed by considering longer chain.}\label{fig:3.3}
\end{figure*}
%

The equivalence no longer holds for the 2D system, since the SOI considered in Hamiltonian \eqref{H} is homogeneous throughout the system, whereas the spiral magnetic ordering is limited only to adatom chain. For $N_y > 1$ the topologically non-trivial region becomes also SOI dependent. In Fig.~\ref{fig:3.3} we show the lowest single-particle energy states as functions of the considered model parameters. Note the transitions of the lowest energy states from finite to zero values when the gaps close and reopen. When $L = N_x$, the occurence of zero energy states coincides with the negative Majorana number, which we calculate using the algorithm developed in Ref.~\cite{Wimmer12}. Note that in some cases the Majorana number is negative, but the minigap is closed, thus in these cases the lowest energy states are not MZEM.

\subsection{Wave functions}\label{sec:IIIC}
\begin{figure}[bht]
\centering
\includegraphics[width=0.5\textwidth]{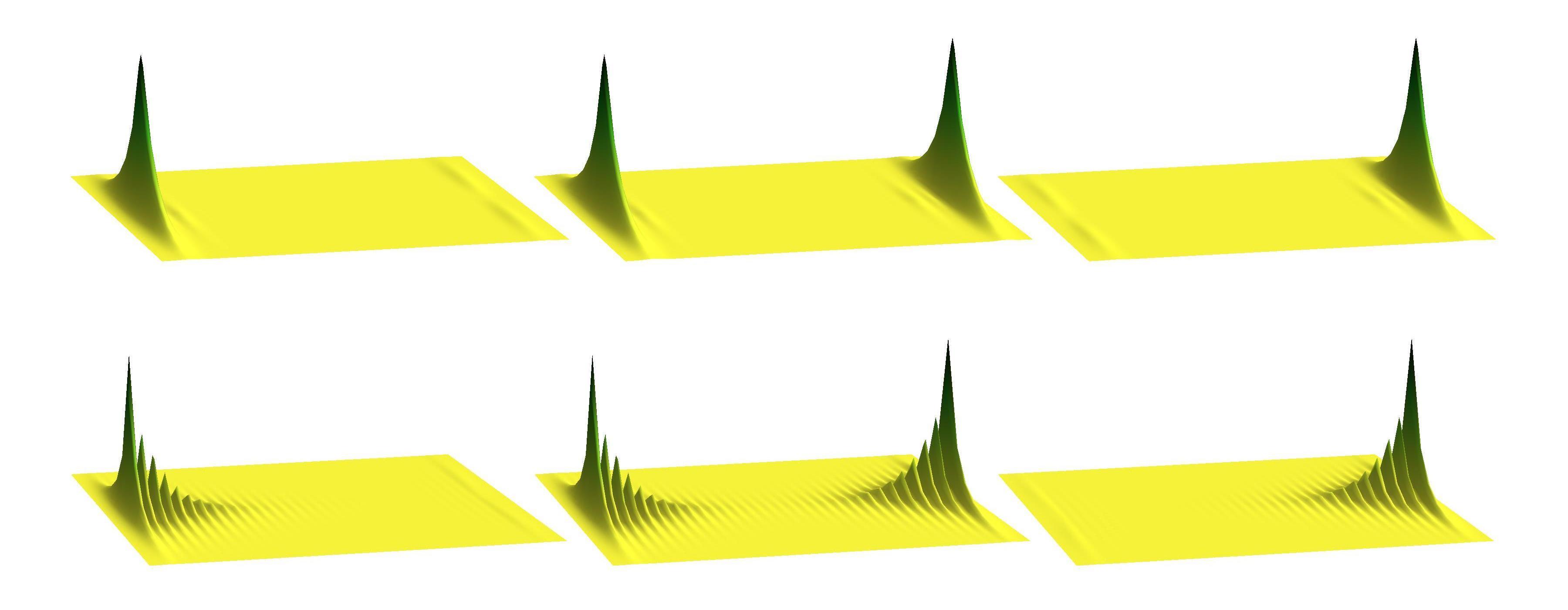} 
\caption{Majorana zero energy states: middle panels show $|\Psi|^2$ and left (right) panels linear combinations of particle and hole coefficients $\gamma_L$ ($\gamma_R$) as given by eq.~\eqref{gamma} for the cases $N_x = 200$, $L = 180$, $N_y = 15$, $\alpha = \Delta = 0.1$, $\mu = 4$ and $J = 0.7$ (upper panels) and $J = 1.1$ (lower panels).}\label{fig:3.4}
\end{figure}

To study the zero energy modes in more detail we analyze their wave-functions. The wave-function $|\Psi|^2$ is given as
\begin{eqnarray}
|\Psi(i)|^2 = \sum_s |u(i,s)|^2 + |v(i,s)|^2
\end{eqnarray}
where $u,v$ are the particle and hole coefficients, respectively and the sum runs over spin $s = \uparrow, \downarrow$. Majorana bound states appear at the end of the FM chain, as can be seen in Fig.~\ref{fig:3.4}, middle panels. In case of Majorana states the particle and hole coefficients are related due to the particle-hole symmetry~\cite{Leijnse12} and the two end state peaks are separated by taking the following linear combinations,
\begin{eqnarray}\label{gamma}
|\gamma_{L,R}(i)| &=& |u(i,\uparrow) \pm v(i, \uparrow)^*|^2 + \nonumber \\
&+& |u(i,\downarrow) \mp v(i, \downarrow)^*|^2,
\end{eqnarray}
as shown in left and right panels of Fig.~\ref{fig:3.4}. This way the separate end states are obtained directly from the particle and hole coefficients calculated from the BdG Hamiltonian \eqref{calH}.

\begin{figure}[bht]
\centering
\includegraphics[width=0.45\textwidth]{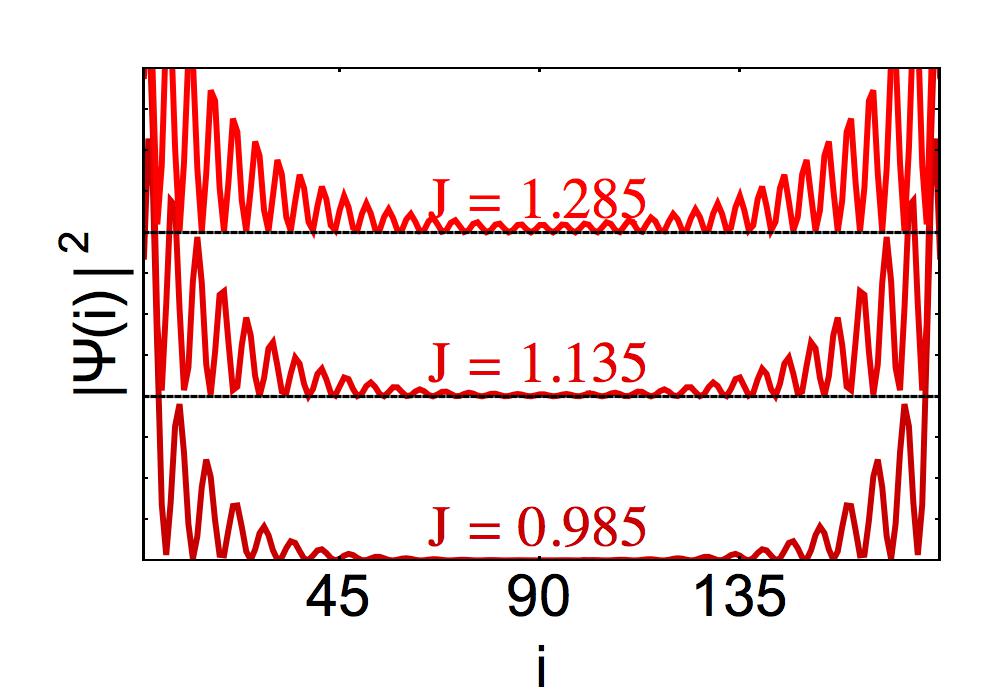}
\includegraphics[width=0.45\textwidth]{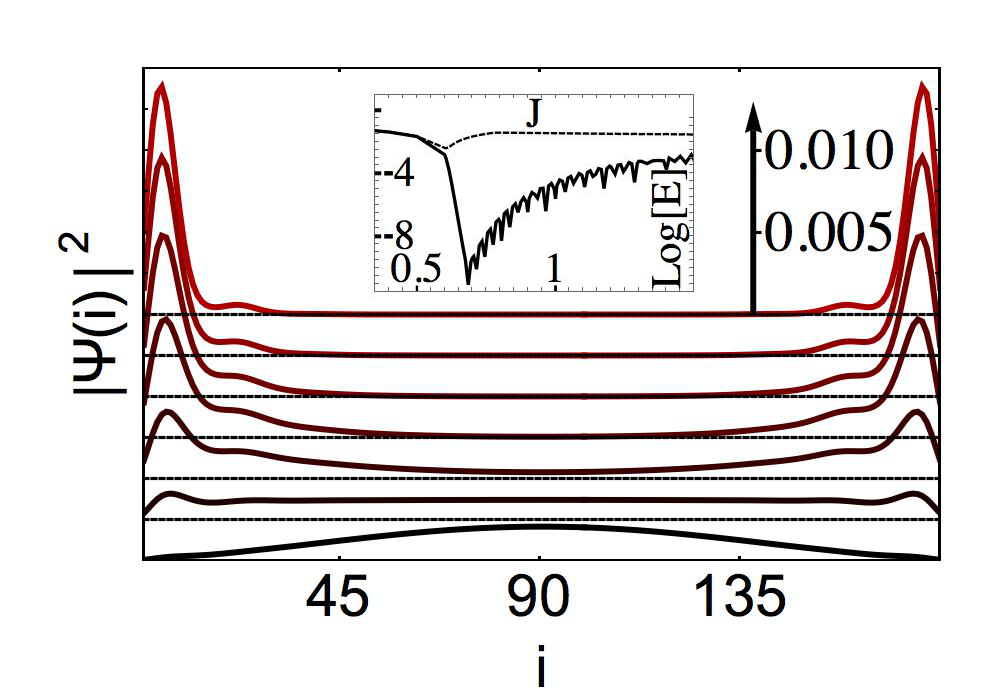}
\caption{Evolution of the MZEM from YSR state: both panels show $|\Psi|^2$ on the adatom sites and different states are displaced for better presentation, indicated also by thin dashed lines. YSR state is the lowest, black curve in lower panel, for which the exchange coupling $J = 0.605$. Second curve from bottom has $J = 0.635$ and between the following neighbouring curves there is difference in the exchange coupling of $0.01$. $J$ grows from bottom to top, so that the top curve with localized end state has $J = 0.685$. Upper panel shows delocalization of end states for increased exchange couplings. The y-axis scale is the same in upper and lower panel and given by the arrow. The logarithms of the two lowest energy states as a function of exchange coupling are shown in the inset. Note that the ground state energy (thick) drops close to zero and - due to enhanced overlap of the end states - grows with further increase of the exchange coupling. Other parameters used are $N_x = 200$, $L = 180$, $N_y = 15$, $\alpha = \Delta = 0.1$, $\mu = 4$.}\label{fig:3.5}
\end{figure}

Typical evolution of MZEM as a function of exchange coupling is presented in Fig.~\ref{fig:3.5}, where we show the wave-function $|\Psi(i)|^2$ and limit to $i$ only at the sites of the adatom chain. Note that before the closing of the gap (small exchange coupling) the wave-function is spread all along the chain, corresponding to the YSR state, whereas at the transition to topological phase, the wave-functions become strongly localized at the ends of the chain. Unlike in 1D systems~\cite{Kitaev01, Lutchyn10, Oreg10} the transition here is more gradual. By further increasing the exchange coupling, the decay of the wave-function becomes larger and the energy of the lowest energy state increases (inset in Fig.~\ref{fig:3.5}).

\subsection{Wider chains}\label{sec:IIID}

\begin{figure}[bht]
\centering
\includegraphics[width=0.5\textwidth]{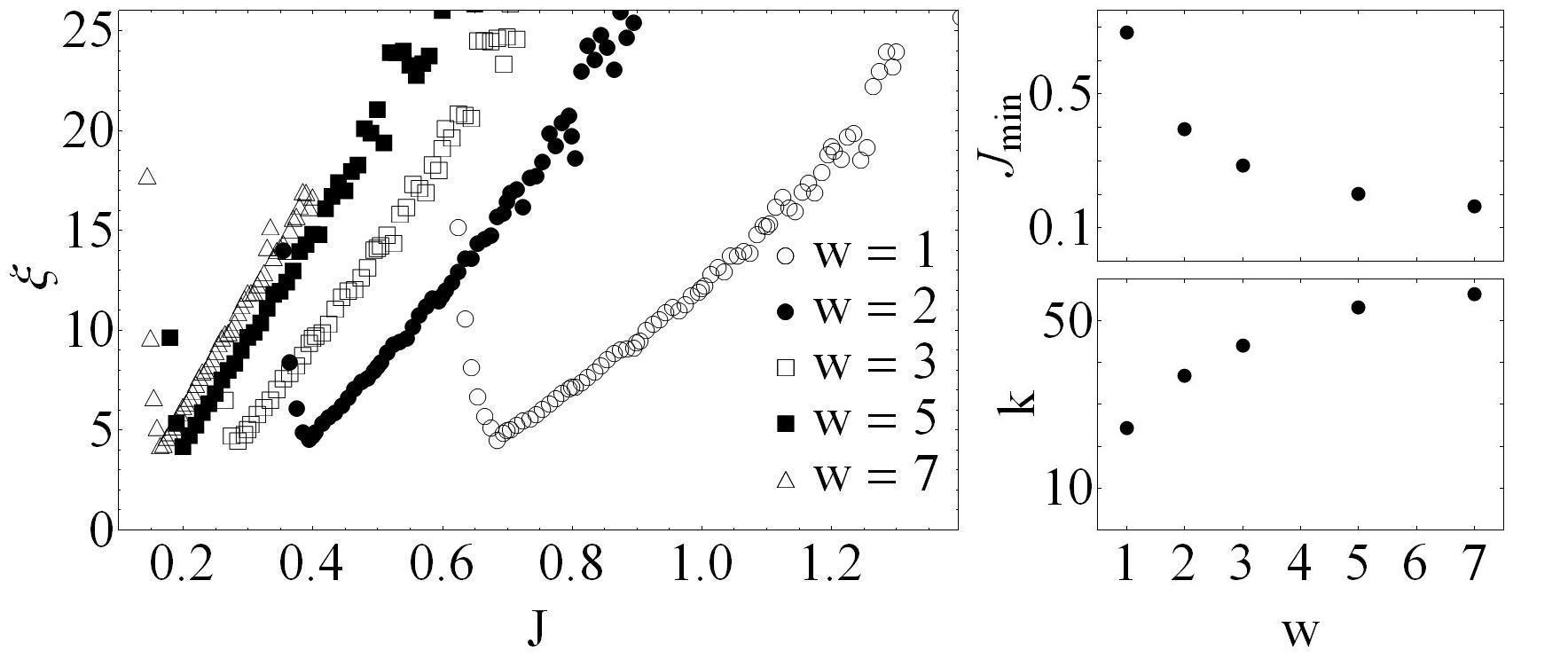}
\caption{Majorana decay lengths as a function of exchange coupling for various chain widths $w$ are shown in left panel. Right panels show the exchange coupling where the minimum of the decay length is reached (upper panel) for different widths. Lower right panel shows the growth of $k$ with increasing chain widths, where $k$ is the slope of the linear fit of $\xi(J) = k J + \xi_0$. System dimensions are $N_x = 200$, $L = 180$, $N_y = 15$ ($N_y = 16$ for $w = 2$) and $\alpha = \Delta = 0.1$.}\label{fig:3.7}
\end{figure}

We study also the wave-function decay length $\xi$, which has been modeled~\cite{Zyuzin13, Nadj-Perge14} as $|\Psi(i)|^2 \sim 1/\sqrt{\xi} e^{-x/\xi}$. Strong localization of MZEM observed in experiment~\cite{Nadj-Perge14} was explained recently by modeling system as a linear chain of Anderson impurities on top of superconductor and solving it analytically with mean-field theory~\cite{Peng15}. As in the theoretical treatment supplementing experiment and experiment itself~\cite{Nadj-Perge14} we find here strongly localized MZEM. Due to the evolution of highly localized state from the YSR state, the decay length first decreases and we note that in this transition region the fitting function poorly describes the actual envelope function. After the transition, the decay length is a linear function of exchange coupling (as can be seen in Fig.~\ref{fig:3.7}) and roughly independent on the chemical potential. With the exception of the transition regions, the decay length shows linear behaviour with negative slope also as a function of the superconducting gap function (effective attractive interaction) for the non self-consistent (self-consistent) treatment of superconductivity.

We consider also cases with various chain widths $w$. Note that the length dependence was studied previously~\cite{Nadj-Perge14} and also various widths were considered but here we systematically study the width effect on Majorana decay lengths and show its implications for experiment. The general observation in this case is that wider chains correspond to a narrow chain ($w = 1$) with increased effective exchange coupling. This can be seen in the Majorana decay lengths in Fig.~\ref{fig:3.7}. The wider the chain, the MZEM appear at smaller exchange coupling, which is also seen in the plot of the exchange coupling where the minimum decay length is shown for different chain widths.  This minimum is the same for all the widths considered and has for the chosen parameters value $\sim 4$. Besides the shift of the decay length minimum also the slope of the decay length as the function of the exchange coupling increases with the increasing chain width.

From these results we see that at fixed exchange coupling the wider chains have  longer decay lengths (in case they are in the topological regime with the MZEM). Thus we suggest that the delocalization of the edge modes for wider adatom chains could be observed in the experiments.
\section{Quantum phase transitions}\label{sec:IV}

First let us review known results for few magnetic impurities. In the case of a single magnetic impurity treated classically there is a first order QPT as a function of the exchange coupling, which was first pointed out by Sakurai~\cite{Sakurai70}. The transition corresponds to a level crossing where the superconductor becomes thermodynamically unstable against spontaneous creation of localized quasiparticle excitation (with spin opposite to the local magnetic field due to impurity), which occurs for both non-self consistent and self-consistent (mean-field) treatment of superconductivity~\cite{Salkola97}. In the latter at the QPT both local superconducting gap function and the local spin density have discontinous jumps as can be seen in Fig. 3 of Ref.~\cite{Salkola97}. A jump in the local superconducting order is seen also in the full quantum treatment~\cite{Sakai93}. In that case the QPT occurs only for antiferromagnetic coupling ($J > 0$)~\cite{Satori92} at $T_K/\Delta \simeq 0.3$, where $T_K$ and $\Delta$ are the Kondo temperature and superconducting gap, respectively. Recently the spectral properties of YSR states at finite temperatures were studied using the numerical renormalization group~\cite{Zitko16}.

The case of two magnetic (classicall) impurities exhibits quantum interference effects, which depend on the interimpurity distance and the relative angle between the impurities~\cite{Morr03}. In contrast to the single impurity case the two impurity system can be driven through multiple QPTs, which may or may not be accompanied by a $\pi$ phase shift of the onsite superconducting gap~\cite{Morr06}. At QPTs the local superconducting gap function changes discontinuously and the spin polarization changes for $\pm 1/2$. The gap function renormalization in the system of two impurities was recently revisited~\cite{Meng15}. Similarly to a pair of magnetic impurities also ordered magnetic chains undergo a series of first order QPTs as the exchange interaction of magnetic moments is increased~\cite{Balatsky06, Sacramento07a, Sacramento10}. These phase transitions are induced by numerous level crossings between the ingap states and can be observed in changes of total magnetization of the electron spin density, local spin density, local density of states, various quantum information measures and the local order parameter. In this work we focus on the latter. 

\begin{figure}[bht]
\centering
\includegraphics[width=0.3\textwidth]{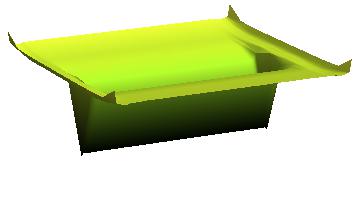}
\caption{Typical superconducting gap function after convergence has a finite positive value away from adatom sites and a drop below the adatom chain, where the value of $\Delta_i$ can be either positive or negative. In the latter case system has undergone a QPT. For the case shown $N_x > L$.}\label{fig:4.1}
\end{figure}

In this section we consider only self-consistent treatment of superconductivity. The typical superconducting gap function profile after convergence is shown in Fig.~\ref{fig:4.1}. It is usually reduced to a small absolute value, which can be either positive or negative on the magnetic adatom sites and assumes a finite positive value away from them. Similar superconducting gap profile was reported also in the study of spiral magnetic chain on top of an s-wave superconducting surface~\cite{Reis14}, although in that work the QPTs were not discussed.
 
To give a more detailed analysis we calculate two average values of the gap function, first the average over all sites in the system, $\Delta$, given as
\begin{equation} \label{D}
\Delta = 1/N \sum_i \Delta_i, %
\end{equation}
and second the average over the adatom sites, $\Delta_{\mathrm{Ad}}$, calculated as
\begin{equation} \label{DImp}
\Delta_{\mathrm{Ad}} = 1/L \sum_{i \in {\mathrm{Ad}}} \Delta_i.
\end{equation}

The overall average of the gap function depends very little on the exchange coupling when the chemical potential is kept fixed and its maximum value at $\mu = 0$ is roughly proportional to the attractive interaction strength $V$. This value decreases with the increasing chemical potential until it vanishes and the superconductivity is destroyed at the chemical potential $\mu \sim 4$. This behaviour of $\Delta$ is seen also by the $L+1$-st energy state in Fig.~\ref{fig:3.1} in the parameter ranges where the ingap states are away from the superconducting gap. As a function of spin-orbit interaction the gap has the largest values at small chemical potentials, which decrease towards 0 value with increasing $\alpha$. For larger chemical potential the value of the gap at small $\alpha$ diminishes to 0 at $\mu \sim 4$ and grows to some finite value after some critical $\alpha$ which increases with increasing $\mu$.

\begin{figure*}[bht]
\centering
\includegraphics[width=0.75\textwidth]{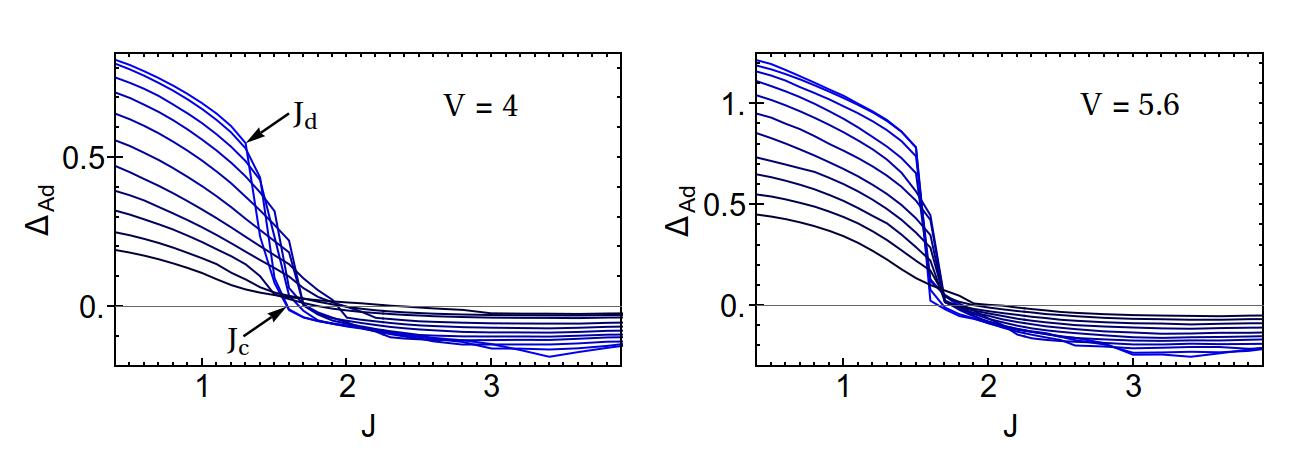}
\caption{Average superconducting gap function at adatom sites for various attractive interactions. The QPTs start in the region $1.3 < J < 1.7$ for $\mu < 4$, where the slope of the gap function changes abruptly. For the exchange coupling line cuts the chemical potential for the highest curve (blue) is $\mu = 0$ and the following curves are in chemical potential steps 0.3. Other parameters used are $L = 90$, $N_y = 11$ and $\alpha = 0.3$.}\label{fig:4.3}
\end{figure*}

Although $\Delta$ itself is very weakly dependent on the exchange coupling, $\Delta_{\mathrm{Ad}}$ shows strikingly different behaviour, as can be seen in Fig.~\ref{fig:4.3}. At small exchange coupling $\Delta_{\mathrm{Ad}} \sim \Delta$ and it slowly decreases with increasing $J$. At some exchange coupling $J_d$ the local gap function suddenly drops. This occurs at $J_d \sim 1.6$ for all differrent chemical potentials and attractive electron interactions considered. The slope change is greater at smaller chemical potentials and it occurs at smaller $J_d$ for smaller chemical potentials and smaller attractive interactions. At higher chemical potentials ($\mu \sim 1.8$ for $V=4$ and $\mu \sim 3$ for $V = 5.6$) $\Delta_{\mathrm{Ad}}$ changes smoothly and there is no abrupt slope change. As exchange coupling is further increased $\Delta_{\mathrm{Ad}}$ passes zero value and becomes negative ($\pi$ shift). This critical exchange coupling $J_c$, where the supercondicting gap at the impurity sites changes sign is between $1.6 < J_c < 2$ for wide range of chemical potentials and attractive electron interactions. By increasing the attractive electron interaction the points $J_d$ and $J_c$ move closer. The $\pi$ shift is also observed by changing the SOI. The critical SOI where the transition occurs dependends on the chemical potential and is for $\mu \sim 0$ at some finite value, which decreases to 0 at $\mu \sim 4$. We observe no $\pi$ shift as a function of attractive interaction $V$.

\begin{figure}[bht]
\centering
\includegraphics[width=0.375\textwidth]{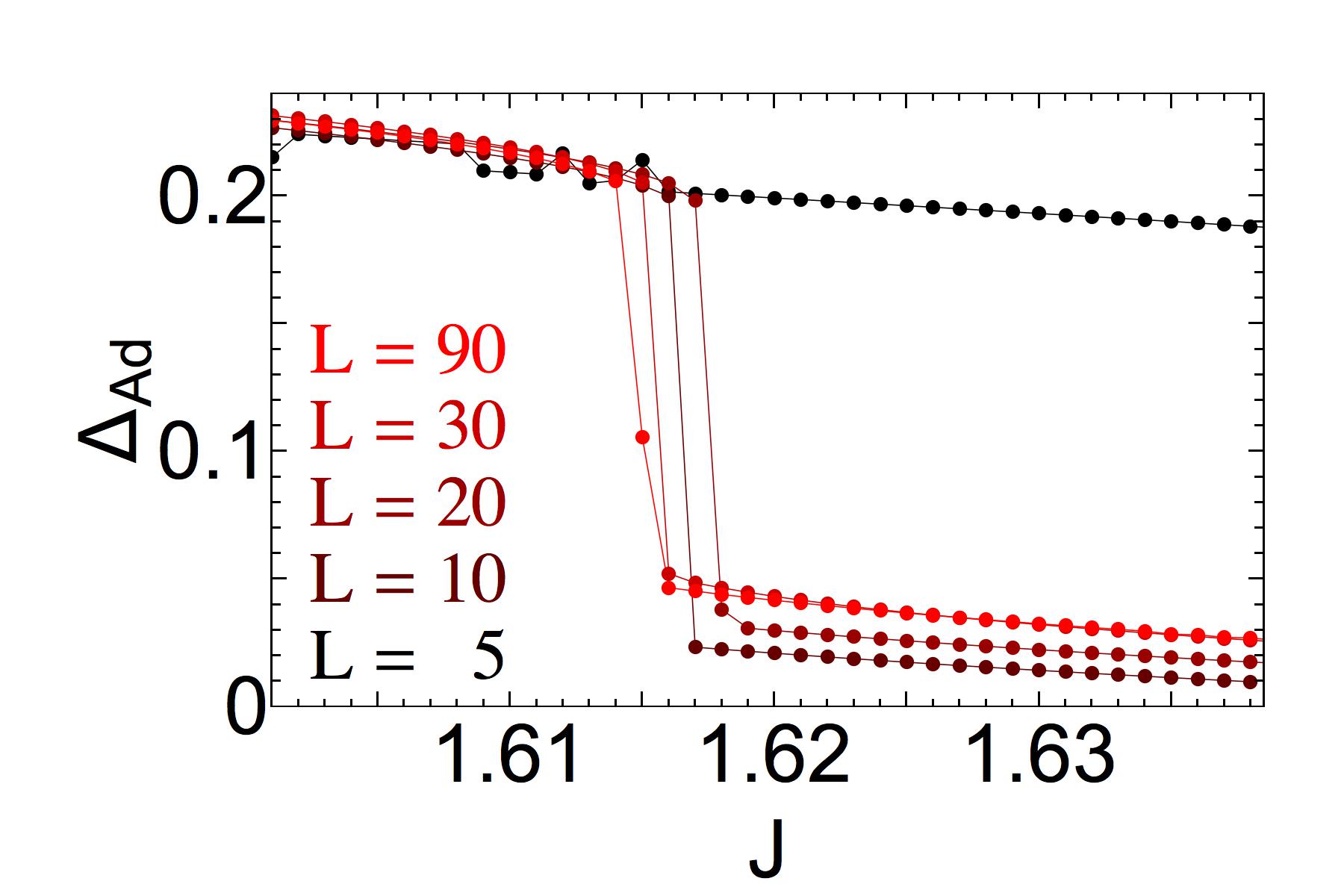}
\caption{Average superconducting gap function at adatom sites as a function of exchange coupling for various chain lengths $L$. $\Delta_{\mathrm{Ad}}$ has a drop where Majorana number changes sign (not shown). Lines are guide to the eyes. The parameters used are $N_y = 11$, $\mu = 1$, $\alpha = 0.3$ and $V = 4$.}\label{fig:4.4}
\end{figure}

Closer inspection of the region in vicinity of $J_d$ reveals that the superconducting gap function on the adatom sites has a finite drop as shown in Fig.~\ref{fig:4.4}. This drop is accompanied with the single particle gap closing and the occurence of MZEM. Consistently the Majorana number calculated using \eqref{M} becomes negative. Remarkably already the case of $L = 10$ gives both a change in Majorana number and a drop in the superconducting gap, which happens practically at the same value of exchange coupling as for longer chains. However, the lowest single particle energy is in this case $\sim 0.03$. Note that even shorter chain of length $L = 5$ does not show a drop in the local superconducting gap function, however the Majorana number also for this case changes sign (at $J = 1.628$). The size of the drop of superconducting gap is a bit larger for smaller system sizes and is the same for longer chains.

The coincidence of the drop in local superconducting gap function and the change of sign in Majorana number is perfect in the whole range of chemical potentials and for various attractive interactions considered where MZEM occur. As can be seen from Fig.~\ref{fig:3.2} this happens around chemical potential $\mu \sim 1$. Thus around $\mu \sim 0$ there are no MZEM, but as soon as they appear we find $J_{TPT} = J_d$. By increasing chemical potential the abrupt decrease of adatom gap function becomes smooth and the chemical potential where this occurs coincides with the disapperance of MZEM in the phase diagram. Since $J_c > J_d$, the superconducting gap function on the adatom sites changes sign already within the area of MZEM, thus the latter exist both at $\Delta_{\mathrm{Ad}} > 0$ and $\Delta_{\mathrm{Ad}} < 0$.

\section{Discussion and Conclusions}\label{sec:V}

The self-consistent treatment of superconductivity was already addressed in systems of magnetic chains on top of conventional superconductors both in the context of QPTs~\cite{Sacramento07b, Sacramento10} and MZEM~\cite{Nadj-Perge13, Reis14}. In the latter it was shown that qualitatively both non self-consistent and self-consistent treatment of superconductivity give very similar results, that is in both treatments MZEM are found in the phase diagrams. Here we confirm such results for a 2D s-wave superconducting surface with a FM chain of adatoms, in the presence of Rashba SOI. We have shown that quantitatively the areas where MZEM occur shift in the phase diagrams, most notable is the shift in the chemical potential (seen in Fig.~\ref{fig:3.2}), where in the case of non self-consistent (self-consistent) superconductivity the MZEM occur at higher (lower) chemical potential, {\it i.e.} $\mu \sim 4$ ($\mu \sim 1$). In the self-consistent treatment the shift is due to the decrease of the superconducting gap with increasing chemical potential (seen in Fig.~\ref{fig:3.1}). Note however that whereas the bulk superconducting gap $\Delta$ is relatively high, the local superconducting gap on the adatom sites $\Delta_{\mathrm{Ad}}$ is small (and even changes sign) in the MZEM region.

By a systematic study of Majorana decay lengths in wide chains (as presented in Fig.~\ref{fig:3.7}) we have shown that wider chains correspond to a narrow chain with increased effective exchange coupling. Due to the linear dependence of the Majorana decay lengths on the exchange coupling that we found, the wider chains have larger decay lengths (at fixed exchange coupling). From this analysis we suggest that STM experiments performed on wider chains should observe greater delocalization of the zero energy modes.

To study the QPTs we have focused on the local superconducting gap function $\Delta_{\mathrm{Ad}}$ (Fig.~\ref{fig:4.3} and \ref{fig:4.4}). As a function of exchange coupling $J$ it decreases gradually with increasing $J$ until a certain value $J_d$, where there is a drop, which is already present at short adatom chain $L \sim 10$ and is only weakly dependent on its length. By further increasing the exchange interaction the superconducting gap $\Delta_{\mathrm{Ad}}$ changes sign ($\pi$ shift). In this work we have shown that the value $J_d$ coincides with the occurence of MZEM at the ends of the chain, thus there is a connection between the topological phase transition and the quantum phase transition. It is interesting to note that there is an analogy between the coincidence of a $\pi$ drop in superconducting gap at the site of single impurity and the QPT and the coincidence of the topological phase transition and the drop of local superconducting gap function at the adatom chain.

T. \v C. would like to thank R. \v Zitko, P. Kotetes, I. Sega, V. Susi\v c and R. Mondaini for fruitful discussions and acknowledge the support by Slovenian ARRS Grant No. P1-0044. Partial support from FCT through grants PEst-OE/FIS/UI0091/2014 and UID/CTM/04540/2013 is also acknowledged.
\bibliographystyle{apsrev4-1}
\bibliography{./bibliography}
\end{document}